\documentclass[11pt]{article}
\usepackage{moriond2000,epsfig}

\def\Journal#1#2#3#4{{#1} {\bf #2}, #3 (#4)}

\def\be{\begin{equation}}
\def\ee{\end{equation}}
\def\bea{\begin{eqnarray}}
\def\eea{\end{eqnarray}}

\begin{document}
\vspace*{4cm}
\title{PROBING THE GALACTIC CENTER BY GRAVITATIONAL LENSING}

\author{S. CAPOZZIELLO$^{(1)}$, G. IOVANE$^{(1)}$, G. LAMBIASE$^{(1)}$,
D.F. TORRES$^{(2)}$ }

\address{$^{(1)}$Dipartimento di Scienze Fisiche ``E.R. Caianiello'',
Universit\`{a} di Salerno \\
 Via S. Allende, Baronissi (SA) I-84081, Italy.\\
$^{(2)}$Instituto Argentino de Radioastronom\'ia, C.C.5, 1894 Villa Elisa,\\
Buenos Aires, Argentina.}

\maketitle\abstracts{The  nature of Galactic Center
could be probed by  lensing experiments
capable of testing the spatial and velocity distributions of stars nearby and
beyond it. Several hypotheses are possible (e.g. massive neutrino
condensation,  boson star) which avoid the shortcomings of the
  supermassive black hole model.}

\section{Introduction}
 Several observational campaigns  have identified
the center of our Galaxy with the supermassive compact dark object
Sagittarius A$^*$ (Sgr A$^*$) which is an extremely loud radio
source \cite{ghez}. Detailed information comes from dynamics of
stars moving in the gravitational field of such a central object.
The statistical properties of spatial and kinematical
distributions are of particular interest: Using them, it is
possible to establish the mass and the size of the object which
are $(2.61\pm 0.76)\times 10^6 M_{\odot}$ concentrated within a
radius of 0.016 $pc$ (about 30 $lds$) \cite {genzel96}. From this
data, it is possible to state that a supermassive compact dark
object is present at the  Galactic Center and, furthermore, it is
revealed by the motion of stars moving within a projected distance
of less than 0.01 $pc$ from the radio source Sgr A$^*$ at
projected velocities in excess of 1000 $km/s$. Furthermore, a
large and coherent counter--rotation, expecially of the
early--type stars, is revealed. Observations of stellar winds
nearby Sgr A$^*$ give a mass accretion rate of $dM/dt=6\times
10^{-6}M_{\odot}yr^{-1}$. Hence, the dark mass must have a density
$\sim 10^9 M_{\odot}pc^{-3}$ or greater and a mass--to--luminosity
ratio of at least $100M_{\odot}/L_{\odot}$. The result is that the
central dark mass is statistically very significant $(\sim
6-8\sigma)$ and cannot be removed even if a highly anisotropic
stellar velocity dispersion is assumed. As a first conclusion,
several authors state that, in the Galactic Center, there is
either a single supermassive black hole or a very compact cluster
of stellar-size black holes. Due to the above mentioned mass
accretion rate, if Sgr A$^*$ is a supermassive black hole, its
luminosity should be more than $10^{40}erg\,s^{-1}$. On the
contrary, observations give a bolometric luminosity of
$10^{37}erg\,s^{-1}$. This discrepancy is the so--called
``blackness problem'' which has led to the notion of a ``black
hole on starvation'' at the  Galactic Center. Besides, the most
recent observations probe the gravitational potential at a radius
larger than $4\times 10^{4}$ Schwarzschild radii of a black hole
of mass $2.6\times 10^{6}M_{\odot}$ \cite{ghez} so that the
supermassive black hole hypothesis
 is far from being conclusive.
On the other hand, stability criteria rule out the hypothesis of a very
compact stellar cluster in Sgr A$^{*}$. In fact,
detailed calculations of evaporation and collision mechanisms give maximal
lifetimes of the order of $10^8$ years which are much shorter than the
estimated age of the Galaxy.
Recently, other viable  alternative models for the
 Galactic Center (and  the center of
several other galaxies) has been proposed. Essentially, the
authors wonder if nonbaryonic condensations, given by massive
neutrinos (or other fermions as gravitinos), or massive bosons
could account for dynamics and size of Sgr A$^*$, without
considering the supermassive black hole
\cite{viollier}$^,$\cite{iovane}$^,$\cite{torres}. The main
ingredient of such proposals is that  nonbaryonic matter interacts
gravitationally forming a supermassive ball in which the
degeneracy pressure of fermions or Heisenberg uncertainty
principle for bosons balance their self--gravity. Both mechanisms,
also if in a completely different way, prevent from gravitational
collapse. Such nonbaryonic condensations could have formed in the
early epochs during a first--order gravitational phase transition.

\section{Sgr A$^*$ as a neutrino star}
Various experiments are today running to search for neutrino
oscillations. It is very likely that exact predictions for
$\nu_{\mu}-\nu_{tau}$ and $\nu_{\mu}-\nu_{\tau}$ oscillations will
be soon available. From all this bulk of data, it is possible to
infer reasonable values of mass for $\nu_{e}$, $\nu_{\mu}$, and
$\nu_{\tau}$. For our purposes, we are particularly interested in
fermions which masses range between 10 and 25 keV$/c^{2}$. This
choice allows the formation of supermassive degenerate objects
\cite{viollier} (from $10^6 M_{\odot}$ to $10^9 M_{\odot}$). with
a the large amount of radio emission. The theory of heavy neutrino
condensates, bound by gravity, can be easily sketched
\cite{viollier}$^,$\cite{iovane} by a Thomas--Fermi model for
fermions. We can set the Fermi energy $E_{F}$ equal to the
gravitational potential which binds the system, that is
${\displaystyle \frac{\hbar^2 k_{F}^{2}(r)}{2
m_{\nu}}-m_{\nu}\Phi(r)=E_{F} =-m_{\nu}\Phi(r_{0}), }$ where
$\Phi(r)$ is the gravitational potential, $k_{F}$ is the Fermi
wave number and $\Phi(r_0)$ is a constant chosen to cancel the
gravitational potential for vanishing neutrino density. The length
$r_{0}$ is the estimated size of the condensation. If we take into
account a degenerate Fermi gas, we get
$k_{F}(r)=\left(6\pi^{2}n_{\nu}(r)/g_{\nu}\right)^{1/3},$ where
$n_{\nu}(r)$ is the neutrino number density and we are assuming
that it is the same for neutrinos and antineutrinos within the
condensation. The number $g_{\nu}$ is the spin degeneracy factor.
Immediately we see that the number density is a function of the
gravitational potential,  i.e. $n_{\nu}=f(\Phi),$ and the model is
specified by it. The gravitational potential will obey a Poisson
equation where neutrinos (and antineutrinos) are the source term:
$
\triangle\Phi=-4\pi Gm_{\nu}n_{\nu}.
$
We can assume
 the spherical symmetry and define the variable
$u=r[\Phi(r)-\Phi(r_{0})]$ then the Poisson equation reduces to the radial
Lan\'e--Emden differential equation
\begin{equation}  \label{n5}
\frac{d^2 u}{dr^2}= -\left(\frac{4\sqrt{2}m_{\nu}^{4}G g_{\nu}}{3\pi\hbar^{3}
}\right) \frac{u^{3/2}}{\sqrt{r}}\,,
\end{equation}
with polytropic index $n=3/2$. This equation is equivalent to the
Thomas--Fermi differential equation of atomic physics, except for the minus
sign that is due to the gravitational attraction of the neutrinos as opposed
to the electrostatic repulsion between the electrons. If $M_{B}$ is the mass
of the baryonic star internal to the condensation, the natural boundary
conditions are $u(0)=GM_{B}\,,u(r_{0})=0.$
The general solution \cite{viollier} of (\ref{n5})
 has scaling properties and it is able
to reproduce the observations. It  well fits the observations
toward the  Galactic Center which estimate a massive object of
$M=(2.6\pm 0.7)\times 10^{6}M_{\odot}$ which dominates the
gravitational potential in the inner ($\leq 0.5$pc) region of the
bulge. In summary, a degenerate neutrino star of mass $M=2.6\times
10^{6}M_{\odot}$, consisting of neutrinos with masses  $ m\geq
12.0$ keV$/c^{2}$ for $g_{\nu}=4$, or $m\geq 14.3$ keV$/c^{2}$ for
$ g_{\nu}=2$, does not contradict the observations. Considering a
standard accretion disk, the data are in agreement with the model
if Sgr A$^*$ is a neutrino star with radius $R=30.3$ ld ($\sim
10^5$ Schwarzschild radii) and mass $M=2.6\times 10^{6}M_{\odot}$
with a luminosity $L\sim 10^{37}$erg sec$ ^{-1}$. Similar results
hold also for the dark object ($M\sim 3\times 10^{9}M_{\odot} $)
inside the center of M87. Assuming the existence of such a
neutrino condensate in the Galactic Center, it could act as a
spherical lens for the stars behind so that their apparent
velocities will be larger than in reality. Comparing this effects
with the proper motion of the stars of the cluster near Sgr A$^*$,
exact determinations of the physical parameters of the neutrino
ball could be possible. In this case, gravitational lensing,
always used to investigate baryonic objects, could result useful
in order to detect a nonbaryonic compact object. Furthermore,
since the astrophysical features of the object in Sgr A$^*$ are
quite well known \cite{genzel96}, accurate observations by lensing
could contribute to the exact determination of particle
constituents which could be, for example, neutrinos or gravitinos.
Our heavy neutrino ball, being massive, extended and transparent,
can be actually considered as a magnifying glass for stars moving
behind it. If an observer is on Earth and he is looking at the
Galactic Center, he should appreciate a difference in the motion
of stars since lensed stars and non-lensed stars should have
different projected velocity distributions. In other words,
depending on the line of sight (toward the ball or outside the
ball) it should be possible to correct or not the projected
velocities by a gravitational lensing contribution and try to
explain the bimodal distribution actually observed
\cite{ghez}$^,$\cite{genzel96}. Detailed calculations in this
sense are given in Capozziello \& Iovane 1999 \cite{iovane} where
the spatial and kinematical distributions of the stars nearby the
Galactic Center are reproduced by using the
 neutrino condensate as a thick lens (a sort of magnifying glass).

\section{Sgr A$^*$ as a boson star}
Also a gravitationally--bound boson condensate could explain
the supermassive object in Sgr A$^*$.
In general, the theory of a boson star can be constructed starting from the
Lagrangian density of a massive complex self-gravitating scalar
field (taking  $\hbar=c=1$)
\be
{\cal L} = \frac {1}{2} \sqrt{\mid g \mid} \left[
  \frac {m_{{\rm Pl}}^2}{8\pi} R + \partial_\mu \psi^\ast \partial^\mu \psi
- U(|\psi |^2) \right ] \; , \label{lagr} \ee
where $R$ is the
scalar of curvature, $|g|$ the modulus of the determinant
of the metric $g_{\mu \nu}$, and $\psi$ is a {\em complex}
scalar field with potential $U$.
Since the potential is a function
of the squared  modulus of the field, we obtain a
global $U(1)$ symmetry. This symmetry
is related with the conserved number of
particles \cite{torres}. The  form of the potential
gives a {\it  mini-boson}, a {\it boson}, or a
{\it soliton} stars.
Conventionally,   it is given by
${\displaystyle
U = m^2 |\psi |^2 + \frac{\lambda }{2} |\psi |^4}$
where
$m$ is the scalar mass and $\lambda$ a dimensionless coupling constant.
Mini-boson stars are
 spherically symmetric equilibrium configurations with
$\lambda =0 $. Boson stars,
on the contrary, have a non-null value
of $\lambda$.
Soliton  stars are
non-topological solutions with a finite mass,
confined in a region of space, and non-dispersive.
They are given by a potential of the form
${\displaystyle
U=m^2 |\psi|^2 \left(1 - \frac{|\psi|^2}{\Phi_0^2}\right)^2},$
where $\Phi_0$ is a constant.
Let us see, now, which are the parameters of the
different scalar objects
which may reproduce the features of the
central object in our Galaxy.
We are  looking for
a mass of $(2.6 \pm 0.76) \times 10^6 M_\odot$,
a radius of the accretion disk of 0.016 pc$\sim$30 light days,
and a luminosity of $10^{37}$ erg s$^{-1}$.
An interesting fact is  that,
for all the above scalar objects, the radius is always related
with the mass in the same
way \cite{torres}: $M=m_{\rm Pl}^2 R$, where $m_{\rm PL}$ is the Planck mass.
In the case of  Galactic Center, it is clear that the main parameter
is the mass and not the radius.
 In the scalar star models, from the given central mass,
the radius we obtain for the star is comparable to that of the
horizon ($R=m_{\rm Pl}^{-2} \times 2.61 \times 10^6 M_\odot=3.9
\times 10^{11}cm$). In other words, we expect an object which
extends about 10 solar radius from the center and which is
singularity free. Due to this intrinsic feature, it is impossible
to use gravitational lensing as above \cite{iovane}: There the
presence of an ``extended" supermassive neutrino condensation in
Sgr A$^*$ allows to distinguish the stars `nearby' and `behind'
the object which effectively acts as a thick spherical lens. The
star spatial positions and projected kinematics have a bimodal
distribution depending on the line of sight (toward the ball or
outside the ball). In the boson condensation case, we have a
pointlike lens (with respect to an observer on Earth) and, also if
the boson star is ``transparent" \cite{DS}, we cannot expect any
bimodal distribution of stars behind it. However, due to the
extremely large mass of the object, a standard gravitational lens
analysis fails, since {\it strong} lensing effects have to be
considered. The observed bimodal distributions in space and
velocities \cite{ghez}$^,$\cite{genzel96} have to be ascribed to
intrinsic, in some sense `genetic', effects. The question is now
for which values of the parameters, we can obtain a scalar object
of such a mass. For the case of mini-boson star, we need an
extremely light boson $m=5.08 \times 10^{-17}$eV/c$^2$. This is
the only parameter of this model and so it is unequivocally fixed.
For boson stars, we get
$
m{\rm [GeV]} = 7.9 \times 10^{-4}(\lambda/4 \pi)^{1/4}.$
It is possible to fulfill the previous
relationship, for instance, with a  boson of about 1 MeV
and $\lambda =1$. In the case of a
non-topological soliton, we obtain
$
m{\rm [GeV]} = 7.6 \times 10^{12}{\rm GeV}^3/(\Phi_0^2 {\rm [GeV]}^2 ).
$
If the parameter $\Phi_0$ is
of the order of boson mass, we need very heavy bosons:
$m = 1.2 \times 10^{4}$GeV/c$^2$.
Based only on the
constraints imposed by the above  mass--radius relationship,
we may conclude that:
$i)$ if the boson mass is comparable
to the expected
Higgs mass (hundreds of GeV),
then the  Galactic Center could be a non-topological soliton star;
$ii)$ an intermediate mass boson could be enough to produce a heavy object
in the form of a boson star;
$iii)$ a mini-boson star
needs the existence of an ultra-light boson.
If boson stars really exist, they should be the remnants of first-order
gravitational phase transitions and their mass should be ruled by the
epochs when they decoupled from the cosmological background. The Higgs
particle, besides its leading role in inflationary models, should
be the best and natural candidate as constituent of a boson condensation,
if the phase transition occurred in early epochs.
A boson condensate should be considered as a sort of topological
defect. In this case, Sgr A$^*$ should be a soliton star.
If soft phase-transitions took place during cosmological evolution
(e.g. soft inflationary events),
the leading particles could have been intermediate mass bosons and so
our supermassive object should be a genuine boson star.
If the phase transitions are very recent, the ultra-light bosons could belong
to the Goldstone sector giving rise to mini-boson stars.
In literature, we can find several examples of particles capable to fit
the issues of boson stars but the ultimate answer is left to the
cosmological observations and particle physics experiments.

\vspace{2.mm}

 As it is widely discussed in literature, gravitational lensing
observations or very large baseline interferometry (VLBI) could
give the ``signature" to discriminate among the  models of
Galactic Center present in literature. For example, the
investigation of the  large ``shadow" of the event horizon of the
central object by an observer (we on Earth) should give
information on dynamics and intrinsic structures. Interesting
proposals and simulations in this sense are given in
\cite{falcke}. Besides, the project ARISE (Advanced Radio
Interferometry between Space and Earth) is going to use the
technique of Space VLBI to increase our understanding of black
holes and their environments, by imaging the extreme physical
configurations produced in their proximities by strong
gravitational fields \cite{arise}. From a theoretical point of
view, developments and results in  gravitational lensing in very
strong field regimes will be of extreme importance \cite{DS}$^{,}$
\cite{VELLIS}.

\end{document}